\documentclass[11pt,a4paper]{article}

\usepackage{amsmath}
\usepackage{amsfonts}

\newcommand{\vf}{\varphi}
\newcommand{\eps}{\varepsilon}

\newcommand{\p}{\partial}
\newcommand{\nn}{\nonumber}

\newcommand{\dvf}{\dot\vf} 
 
\newcommand{\dhh}{\dot h} 
\newcommand{\dF}{\dot F} \newcommand{\ddF}{\ddot F}
 
\newcommand{\dpsi}{\dot \psi} \newcommand{\ddpsi}{\ddot \psi}
\newcommand{\dxi}{\dot \xi} \newcommand{\ddxi}{\ddot \xi}

\newcommand{\cF}{\mathcal{F}}\newcommand{\cG}{\mathcal{G}}
\newcommand{\cI}{\mathcal{I}}

\newcommand{\cL}{\mathcal{L}} \newcommand{\cU}{\mathcal{U}}

\newcommand{\bxi}{\bar{\xi}}

\newcommand{\txi}{\tilde{\xi}}

\newcommand{\qub}{\mathbf{q}} 
\newcommand{\vb}{\mathbf{v}} 
\newcommand{\cb}{\mathbf{c}} \newcommand{\pb}{\mathbf{p}}
\newcommand{\betav}{\vec{\beta}} \newcommand{\psiv}{\vec{\psi}}
\begin{document}

\title{{\bf Dilaton-scalar models in context of generalized affine gravity theories: their properties and integrability. }}

\author{E.~Davydov\thanks{davydov@theor.jinr.ru} ~and A.~T.~Filippov \thanks{Alexandre.Filippov@jinr.ru}~ \\
{\small \it {$^+$ Joint Institute for Nuclear Research, Dubna,
Moscow Region RU-141980} }}
 \maketitle

\begin{abstract}
Nowadays it is widely accepted that the evolution  of the universe
was driven by some scalar degrees of freedom both on its early
stage and at present. The corresponding cosmological models
often involve some scalar fields introduced ad hoc. In this paper
we cultivate a different approach, which is based on a derivation
of new scalar degrees of freedom from  fundamental modifications of
Einstein's gravity. In elaboration of our previous work, we here
investigate properties of the dilaton-scalar gravity obtained by
dimensional reductions of a recently proposed affine generalized
gravity theory. We show that these models possess the same symmetries
as related models of GR with ordinary scalar fields.

As a result, for a rather general class of dilaton-scalar gravity
models we construct additional first integrals and formulate
an integral equation well suited for solving by iterations.
\end{abstract}

\section{Introduction}

The cosmological observations of recent decades revealed that the
universe expanded with acceleration on two different stages of
its evolution: in the very beginning and at present time. In
addition, the presence of new sorts of matter, the amount of which significantly exceeds that of the known matter, is also established
very well. The models within the current paradigm of the
Friedmann-–Lemaitre–-Robertson–-Walker cosmology and the Standard
model physics can hardly describe \emph{all} of them.
This probably means that partial improvements of the existing theory
may prove to be insufficient and the entire paradigm should be
changed \cite{Horvath:2008bc}.

For example, one can try to extend the particle physics by adding
inflaton or quint-essence fields, or consider the modified gravity,
or use the paradigm of string/brane world (see, e.g.,  \cite{Linde,
Ratra,Sahni:2004ai,Nojiri:2006ri} as a drop in the ocean of papers
and reviews on the topic).

Incorporating  new scalar fields into the Standard model could
be a technically simplest solution, but this approach lacks a
fundamental basis. Moreover, it can hardly be falsified, since any
disagreement with the observational data can be cured by
adding more fields to better suit the data. Actually, now it
is widely accepted that the string theory, being a fundamental
one, suffers the same problem, with numerous compactifications
which can provide any effective theory.

Among the modifications of gravity there is a very interesting
theory, originally introduced by Weyl, Eddington and Einstein, which has a fundamental distinction from General Relativity. It
establishes the priority of geodesics within non-Riemannian
geometry, while the metric arises at the level of the effective
theory. On the other hand, the constraints on the initial theory
come from the requirement to reproduce the metric theory of GR
gravity. The constraints can be resolved, and the use of a special
two-step Lagrangian formalism allowed  Einstein to formulate the
effective theory with normal curvature term, cosmological constant
and some `dark' matter fields interacting only with gravity. These
fields looked quite abnormal, and since the actual goal at that
time (around 1923) was to unify the gravity and electromagnetism, the theory was not investigated in detail, soon abandoned and completely forgotten.

Nowadays, the presence of dark fields is the fact of our life, and
such a  natural, almost inevitable, presence of the dynamical cosmological constant is welcomed very much. Indeed, one of the intriguing questions of the dark energy problem is: why does it look so much like $\Lambda$-term \cite{Sahni:1999gb,Sahni:2006pa,Riess:2006fw,WoodVasey:2007jb},
while the origin of a so small $\Lambda$-term is very unlikely in the
framework of GR and Standard model. Thus it looks quite natural to re-establish investigations of the affine generalization of GR
\cite{Filippov:2010bs,ATFp}, which can be called the `affine'
gravity.

One of the basic features of the affine gravity is the presence of
a massive (normal or tachyonic) vector field. In cosmology, the
scalar fields are usually considered, but recently \cite{ATFp} we
revealed the duality between this vector field (vecton) and the
corresponding scalar field (scalaron) in case of reductions to
two-dimensional theory. This allows us to investigate the basic
properties of practically interesting cosmological and static
solutions using the same approach as for gravity theories with
scalar fields. In fact the affine gravity produces a scalaron
field which significantly differs from the standard scalar field
in GR, thus compelling us to use a significantly generalized
description of the dilaton-scalar gravity (DSG). And the obtained
results can be applied even outside the context of affine gravity.
For instance, the dimensional reductions of Yang--Mills fields
\cite{Manton,Volkov:1998cc5, Gal'tsov:1998af,Volkov:2001tb} also
provide rather special scalar degrees of freedom.

On the other hand, DGS theories constructed for the
cosmological purposes, are usually rather simple. In case of
affine gravity theory, the corresponding effective DGS model, that we call the dilaton-scalaron gravity (DSG), is defined rather
strictly. Actually it \emph{looks} quite cumbersome, being
significantly non-linear in dimensions $D>3$ and non-integrable
even in the simplest linear case.  Nevertheless, in this paper we will investigate the basic properties of the DGS model and try to answer the question: is the DSG model \emph{really} complicated?

The paper is organized in the following way. First, we discuss
models which possess some fruitful properties, such as the presence
of linear symmetry, the availability of iterative integration
procedure, the existence of explicit non-trivial solutions. So we
specify the set of simplest models within the class of DGS, mostly
related to dimensional reductions. Then we show that DSG
belongs to this set, despite its specific non-linearity.
We thus conclude that affine gravity theory, being quite awkward at
first sight, shares with common scalar field models in GR
some useful properties. In Appendix we present some
extra results which were derived while investigating the DGS
gravity but can be applied to a  rather general class of dynamical
systems appearing in theories with vanishing Hamiltonian constraint.

\section{Dilaton-gravity theory with scalar fields}

A fairly general higher-dimensional field theories can be reduced to
two-dimensional effective DGS models by taking into account their
space-time symmetries. We specify the following generic action
\begin{equation}\label{Laffine}
    \cL_{\textrm{eff}}^{(2)}=\sqrt{-g}\left[\vf
    R(g)+W(\vf)(\nabla_\mu\vf)^2+
    Z_{ij}(\vf;\psi)\nabla_\mu \psi_i \nabla^\mu
    \psi_j+X(\vf;\psi)\right],\quad \mu=(t,r),
\end{equation}
where $\vf$ is the dilaton field, and $\psi_i$ denote all scalar
fields of any origin, including the geometrical one. Note that in
dimension $D=2$ all the fields are dynamically
equivalent to scalar fields or, in case of the massless vector, to the corresponding field charges.  A practical transformation of the vecton into the scalaron in the dimensionally reduced affine gravity model
  was given in \cite{ATFp}.

The most concrete results for these models can be obtained after
the further reduction to the one-dimensional dynamical systems,
describing static and cosmological states. For this, the
system is parameterized by a single coordinate, $\tau$, which can
be time-like or space-like. The signs of the one-dimensional
Lagrangian components depend on the parametrization, but they
usually may be consumed by the free parameters of the model.

So, the corresponding one-dimensional Lagrangian can be written in
the following rather general form \cite{ATFp,ATF1,ATF1a,ATFr,ATFw}:
\begin{equation}
\label{3EqL}
\cL=-s\left[h^{-1}\dot{h}\dvf+W(\vf)\dvf^2+Z_{ij}(\vf;\psi)
\dpsi_i \dpsi_j\right]+s^{-1}h X(\vf;\psi).
\end{equation}
Here $h$ is a metric function, chosen so that $h>0$ is a
cosmological state, and $h<0$ is a static state.
The presence of the remaining metric gauge degree of freedom $s$,
which is in fact the Lagrangian multiplier, leads to the vanishing Hamiltonian (constraint):
\begin{equation}\label{3EqH}
H=s\left[h^{-1}\dot{h}\dvf+W(\vf)\dvf^2+Z_{ij}(\vf;\psi) \dpsi_i
\dpsi_j\right]+s^{-1}h X(\vf;\psi)=0.
\end{equation}
After imposing this constraint we may fix $s$. For example, $s=1$
is the lightcone gauge. In what follows we often replace the metric $h$ with the new variable $F$ defined by
$\dF=h^{-1}\dot{h}+W(\vf)\dvf\,$. This
reduces the dilaton-metric kinetic term to $\dF\dvf$ and incorporates the Weyl transformation; the potential $X(\vf;\psi)$ accordingly
acquires the factor $\Omega(\vf)=\exp(-\int W(\vf)d\vf)$.

\subsection{Linear symmetries and bilinear integrals in DGS}

Consider now the system~(\ref{3EqL})  with just one scalar field
$\psi$ and the corresponding function, $Z(\vf)$, depending only on
the dilaton variable. In this case, we may use a convenient gauge
$s=1/Z(\vf)$  and the new dilatonic variable $\xi$ defined by
$\dxi\equiv\dvf/Z(\vf)$. Then our DGS dynamical system with the
Hamiltonian constraint looks rather compact:
\begin{equation}\label{3L}
  \cL=\dF\dxi+\dpsi^2-e^F U(\xi,\psi),\quad H=\dF\dxi+\dpsi^2+e^F U(\xi,\psi)=0,\quad \mbox{where} \quad U=\Omega ZX.
\end{equation}
The corresponding Lagrangian system,
\begin{equation}\label{3Eq}
  \ddxi=- e^F U,\quad
  \ddF=-e^F U_\xi ,\quad
  \ddpsi=-e^F U_\psi/2,
\end{equation}
actually contains only two independent equations, while the
remaining one can be derived from the above Hamiltonian constraint
$H=0$. Yet it is more convenient to use all the three equations when
searching for \emph{linear symmetries} of the system.

With this aim, let us introduce the coordinate vector
$q^{i}=(F,\xi,\psi)$ and represent our dynamics in a more general form,
\begin{equation}\label{3Lq}
  \cL=A_{ij}\dot{q}^i\dot{q}^j-\cU(q),\quad H=A_{ij}\dot{q}^i\dot{q}^j+\cU(q)=0,
\end{equation}
where in model (\ref{3L}): $A_{F\xi}=A_{\xi F}=1/2$, $A_{\psi\psi}=1$, and $\cU(q)=e^F U(\xi,\psi)$. The equations are
\begin{equation}
  2A_{ij}\ddot{q}^j=-\p_i\cU(q) \quad \Rightarrow\quad \ddot{q}^k=-\frac12(A^{-1})^{ki}\p_i\cU(q).
\end{equation}
The last system can be contracted  with a linear combination  of the
coordinates and some constants, $(\lambda A_{kj}+B_{kj})q^j+c_k$, where
$B_{kj}$ is an arbitrary antisymmetric constant matrix, $c_k$ is
a constant vector, and $\lambda$ is a constant. Then we have:
\begin{equation}
\begin{split}
  [(\lambda A_{kj}+B_{kj})q^j+c_k]\ddot{q}^k=&\{[(\lambda A_{kj}+B_{kj})q^j+c_k]\dot{q}^k\}\,\dot{}-\lambda A_{kj}\dot{q}^k \dot{q}^j=\\
&=-\frac12[(\lambda A_{kj}+B_{kj})q^j+c_k] (A^{-1})^{ki}\p_i\cU(q).
\end{split}
\end{equation}
Using here constraint (\ref{3Lq}), $H = 0$, we replace
the kinetic term, $A_{kj}\dot{q}^i \dot{q}^j$, by
the potential, $-\cU(q)$, and conclude that the bilinear form,
\begin{equation}\label{eqP}
  P=[(\lambda A_{kj}+B_{kj})q^j+c_k]\dot{q}^k \,,
\end{equation}
is the \emph{integral of motion} if the potential satisfies the linear  partial differential equation:
\begin{equation}\label{eqU}
  \lambda \cU(q)+\frac12[\lambda q^i+(B_{kj}q^j+c_k)(A^{-1})^{ki}]\p_i\cU(q)=0.
\end{equation}

  This theorem is applicable to the general constrained dynamics (gauge system) described by Eqs.(\ref{3Lq}) with any constant matrix  $A_{ij}$ and any function $\cU(q^1,...,q^n)$.\footnote{
  Note that throughout the paper we consider the simplest DGS  with one scalar matter mode. The general system is briefly discussed in Appendix~A and multi-scalar DGS models -- in Appendix~B.}
  For our DGS system with one scalar field, the only nonzero components of the $3\times3$ inverse matrix $A^{-1}$, are $(A^{-1})^{12}=(A^{-1})^{21}=2$, $(A^{-1})^{33}=1$ and the potential has the simple dependence on the metric, $\cU(q)=e^F U(\xi,\psi)$. In addition, in the three-dimensional coordinate space, the antisymmetric matrix  $B_{kj}$ can be parameterized by a vector, $B_{kj}=\epsilon_{kjl}b^l$. It follows that Eq.(\ref{eqU}) splits into  two independent equations for $U(\xi,\psi)$ (the second one expresses the condition of $F$-independence):
 \begin{eqnarray}
 && \left[(\lambda/2+b_3)\xi-b_2\psi+
c_1\right]U_\xi+\frac12\left[\lambda\psi-b_1\xi+c_3\right]U_\psi
+(b_1\psi+c_2+\lambda)U =0 \,,\nn \\
&& b_2 \p_3 U \equiv b_2 U_\psi=(2b_3-\lambda)U \,.
\label{Ucond}
\end{eqnarray}
This equation depends on seven parameters, but $U$ is defined up to
an arbitrary multiplier. Integral (\ref{eqP}), also defined up to a multiplier, depends on the same parameters and now reads:
\begin{equation}\label{PDGS}
  P=\left[(\lambda/2+b_3)\xi-b_2\psi+c_1\right]\dF+
  \left[(\lambda/2-b_3)F+b_1\psi+c_2\right]\dxi+\left[\lambda\psi+b_2 F-b_1\xi+c_3\right]\dpsi \,.
\end{equation}

 Note that when $b_2 \neq 0$ and $(\lambda - 2b_3) \neq 0$ the potential defined by equations (\ref{Ucond}) must be exponential in $\psi$, i.e.,
  $U = u(\xi) \exp [(2 b_3 - \lambda) \psi/b_2]$. Inserting this into the first equation we find that $u(\xi)$ is also exponential (or constant). So the model reduces to the well studied class of multi-exponential DGS theories with the completely integrable subclass of Toda - Liouville theories (see \cite{multi} and references there)
   and we therefore expect to find in the exponential case some integrable systems. Below, we mostly suppose that $b_2 = (\lambda - 2b_3) = 0$ that does not mean ignoring exponential potentials. In fact, supposing that $\lambda = b_i = 0$, it is not difficult to find the general solution of (\ref{Ucond}),
  \begin{equation}
 Y(\xi,\psi)\equiv\ln |U| = f(c_3 \xi - 2 c_1 \psi) + c_4 \xi -2 c_3^{-1}(c_1 c_4 + c_2)\,,
  \end{equation}
  where $f$ is an arbitrary function and $c_4$ --- a new arbitrary constant. Making the simplest choice, $f(x) = c_5 x$ we find that
  $Y = g_1 \xi + g_2 \psi$, with $g_1 \equiv c_3 c_5 + c_4$
  and $g_2 c_3 \equiv c_1 g_1 + c_2$. The potential depends only on the parameters $g_1$, $g_2$ and thus we are free to vary all the parameters $c_i$
  while not changing $Y$.  For example, we can take $c_1 =0$, $2c_2 = -g_2 c_3$ or
   $c_2 =0$, $2g_1 c_1 = -g_2 c_3$ and obtain two independent integrals
   discussed below.

   These examples show that relations between the potential and integrals are in general rather complex. We hope to demonstrate that the construction given by Eqs.(\ref{eqP}) - (\ref{PDGS}) is a powerful device,  but it is not easy to use. In this paper, we only briefly discuss mathematics behind it and concentrate on simple but nontrivial examples. A very close but simpler approach to searching for integrals in DGS models can be found in \cite{ATF13}.

\subsection{Master Integral Equation in DGS}
One may reduce the order of the system~(\ref{3Eq}) by choosing
$\xi$ as the independent variable with $d\tau\equiv
d\xi/\chi(\xi)$. The system of equations can be therefore
simplified if we define $\cF\equiv F-\ln\chi$, $\cG\equiv\psi'$,
and use the combination of constraint~(\ref{3L}) with the first
equation of the system instead of the second one. Then the
system will contain only the first order equations:
\begin{equation}\label{eqxi}
  \chi'=-e^\cF U,\quad \cF'=-\cG^2,\quad \psi'=\cG,\quad (\chi\cG)'=-e^\cF U_\psi/2\,.
\end{equation}

Obviously, the solution for $\psi$ and $\cF$ can be written as formal integrals for $\cG$ and $\cG^2$:
\begin{equation}
    \psi(\xi)=\psi_0+\int_{\xi_0}^\xi\cG(\bxi) d\bxi\equiv
    \cI\{\cG;\xi\},\quad \cF(\xi)=\cF_0-\int_{\xi_0}^\xi\cG^2(\bxi) d\bxi\equiv
\cI\{\cG^2;\xi\}.
\end{equation}
Now use these integrals in the remaining equations for $\chi$ and $\chi\cG$ and integrate them as well:
\begin{eqnarray}
&&
\chi(\xi)-\chi_0=-\int_{\xi_0}^{\xi}\exp\left(\cI\{\cG^2;\bxi\}\right)U(\bxi,\cI\{\cG;\bxi\})d\bxi\equiv
\cI_1\{\cG;\xi\},\\
&&2[\chi(\xi)\cG(\xi)-\chi_0\cG_0]=-\int_{\xi_0}^{\xi}\exp\left(\cI\{\cG^2;\bxi\}\right)U_\psi(\bxi,\cI\{\cG;\bxi\})
    d\bxi\equiv \cI_2\{\cG;\xi\}.
\end{eqnarray}

In the last equation replace $\chi$ with the functional $\cI_1$ and find the relation between $\cG$ and the functionals depending on $\cG$ and the potential $U$:
\begin{equation}\label{MIE}
    \cG(\xi)=\frac{\chi_0\cG_0+\cI_2\{\cG;\xi\}/2}{\chi_0+\cI_1\{\cG;\xi\}}.
\end{equation}
We call this the `Master Integral Equation', or MIE. It can be
used to obtain the approximate solutions for the general
potential. With a clever choice of the trial function for $\cG$
we can obtain a good approximation even with a few iterations.
Although the integrations are in general rather cumbersome,
the computations with MIE may be very effective near singular points
 or in vicinity of a horizon, where the solution can be expressed with the aid of convergent power series expansions (after explicitly
 resolving possible singularities).

\section{Partially integrable DGS models}

The models of gravity with scalar fields often can not be solved
explicitly. Yet in case of just one massive scalar mode one may
expect that deriving an additional first integral and/or applying
iterative approximation procedure will allow to investigate the
most important properties of the solutions. For example, in cosmology
the slow-roll approximation technique provides a good inflation
model easily derived for many non-integrable configurations. Having
this in mind, consider some types of DGS configurations possessing the
inner symmetries, which simplify their analysis.

\subsection{Integrable DGS models}
Let us first explore the completely integrable configurations. They
may arise in case of the exponential potential linear in its
arguments: $U=\exp{(g_1\xi+g_2\psi)}$. It possesses two commuting
linear first integrals (the conditions for existence of
several commuting first integrals can be found in Appendix).
Indeed, such potential suits the condition~(\ref{Ucond}) with the non-vanishing parameters $c_2=-c_1g_1$, or $c_2=-c_3g_2/2$. The two
 independent integrals,
\begin{equation}\label{I2exp}
  P_1=\dF-g_1\dxi,\quad P_2=2\dpsi-g_2\dxi,
\end{equation}
are in involution (commute). Taking into account the
constraint,  we find that the number of the integrals is equal to
the number of variables and thus the system is integrable. This
configuration with linear exponential potential may be related to
the Toda-Liouville systems \cite{ATFw}, or it can be derived from a quite special cylindrical reduction, as we will show below.

Another example comes from the effective $D=2$ massless
scalar-vector configuration
\begin{equation}\label{Lsv}
    \cL=\sqrt{-g^{(2)}}\left[\vf
    R(g^{(2)})+W(\vf)(\nabla_\mu\vf)^2+
    Z(\vf)\nabla_\mu \psi \nabla^\mu
    \psi+Y(\vf)B_{\mu\nu}B^{\mu\nu}+X(\vf)\right],
\end{equation}
where $B_{\mu\nu}=\p_\mu A_\nu-\p_\nu A_\mu$. In  the one-dimensional cosmological or static case, the scalar and vector fields
can be integrated out, leaving as a trace the effective charge of the vector field, $Q_0$, and the integration constant of the scalar field,
$c_0 = Z(\xi) \dxi$. So the dynamics may be described by
the effective Lagrangian supplied with the Hamiltonian constraint:
\begin{equation}\label{Lmassless2}
\cL=-\dF\dxi+e^F U(\xi)+V(\xi), \quad H=\dF\dxi+e^F
U(\xi)+V(\xi)=0\,.
\end{equation}
Here the original kinetic scalar field term contribute into the
effective potential $V$, while the kinetic vector term contribute
into $U$.

To derive an explicit solution of such a model we now need just one
additional integral. Let us use here a sort of an inverse approach:
first choose the suitable model which does possess such an integral,
and then find a corresponding physical theory described by the
model.
First check the general condition~(\ref{eqU}) for the (bi)linear
integral to exist. For the system~(\ref{Lmassless2}) it can be
satisfied only when one of the effective potentials, either $U$ or
$V$, vanishes, because the total potential term,
$e^FU(\xi)+V(\xi)$, does not possess the linear symmetry.
Therefore, the mixed system with scalar and vector fields cannot
have such kind of integral.
With both potentials non-vanishing, there remains a
possibility for a non-linear integral to exist. There is no
general algorithm for  constructing  non-linear integrals but, nevertheless,
 an integral quadratic in momenta was found for a dynamical system of this sort (in \cite{ATF1}, \cite{ATF1a}).

Reproducing the procedure of \cite{ATF1} in our notation, first
rewrite the equation for the metric:
\begin{equation}
\ddF=-e^FU'(\xi)-V'(\xi).
\end{equation}
 For linear potentials, $U(\xi)=g_1\xi+g_3$, $V(\xi)=g_2\xi+g_4$,
 it becomes trivial and can be integrated:
\begin{equation}\label{intR}
\ddF=-g_1 e^F-g_2\,, \qquad
  \frac{1}{2}\dF^2=C-g_1e^F-g_2 F\equiv R(F).
\end{equation}
Here $C$ is the new first integral, quadratic in momenta and
non-linear in the coordinate variable $F$.  Multiplying it by $\dxi$
and expressing $\dF\dxi$ by using the Hamiltonian constraint we
find:
\begin{equation}
  [e^F(g_1\xi+g_3)+g_2\xi+g_4]\dF+2(C-g_1e^F-g_2F)\dxi=0.
\end{equation}
This expression is linear in momenta and may be more convenient to
deal with.

Supposing that $g_1\neq 0$, let us rescale $\xi$, $\txi=g_1\xi+g_3$, and rewrite the above equation as
\begin{equation}
  [(g_1e^F+g_2)\txi+g_4g_1-g_2g_3]\dF+2R(F)\dot{\txi}=0.
\end{equation}
Since $g_1e^F+g_2=-R'(F)$, we find the equation
\begin{equation}
  2R(F)\dot{\txi}-\txi\dot{R}+(g_4g_1-g_2g_3)\dF=0,
\end{equation}
which can be integrated after dividing it by $R^{3/2}$:
\begin{equation}
  2R^{-1/2}(F)\txi-2R^{-1/2}(F_0)\txi_0=(g_2g_3-g_4g_1)\int_{F_0}^F R^{-3/2}(\bar{F})d\bar{F}.
\end{equation}

It is instructive to rewrite the the final expression for $\txi(F)$
in the form:
\begin{equation}\label{intsol}
  \txi= \sqrt{\frac{R(F)}{R(F_0)}} \, \biggl[\txi_0 +
 \frac{1}{2} (g_2g_3-g_4g_1)\sqrt{R(F_0)}  \int_{F_0}^F R^{-3/2}(\bar{F})d\bar{F} \, \biggr].
\end{equation}
With $g_2=0$, the integral in r.h.s. will contain arctan, arctanh
or exponential functions depending on the parameters, otherwise it cannot
be expressed in terms of elementary functions. Let us also note
that the model with $g_2=0$ may describe the spherical dimensional
reduction of the massless scalar and vector configuration~\cite{ATF1,ATF1a}. For example, the Reissner--Nordstr\"{o}m solution in the absence of
the scalar field (when $c_0 = 0$) can be derived from~(\ref{intsol}) when
$g_2=g_4=0$.

We would like to mention that the full integrable model with all
$g_i$ non-vanishing provides a rather rich dynamics as compared to the
Schwarzschild or RN solutions. The investigation of such fairly complex integrable models can significantly improve our understanding of complex dynamics in non-integrable theories with massive fields.

\subsection{DGS and cylindrical reduction}

The general cylindrical reduction is, in fact, much more
complicated than a spherical one. Consider the decomposition of
the line interval in $D=4$:
 \begin{equation}
 ds_4^2 = (g_{ij} + \vf \,\sigma_{mn} \,\vf_i^m \vf_j^n) \, dx^i dx^j +
  2 \vf_{im} \, dx^i dy^m + \vf \, \sigma_{mn} \, dy^m dy^n \, ,
 \label{2.7}
 \end{equation}
  where $i,j = 0,1$, $m,n = 2,3$, all the
 metric coefficients depend only on the $x$-coordinates ($t,r$),
 and $y^m =(\phi, z)$ are coordinates on the two-dimensional
 cylinder (torus). Note that $\vf$ plays the role of a dilaton and
 $\sigma_{mn}$ ($\det \sigma_{mn} = 1$) is the so-called
 $\sigma$-field.

As was shown in \cite{ATFr}, the reduction of the Einstein part of
the four-dimensional
 Lagrangian, $\sqrt{-g^{(4)}} \, R(g^{(4)}) $, can be written as:
 \begin{equation}
 \cL_{\textrm{c}}= \sqrt{-g^{(2)}} \, \biggl[ \vf R(g^{(2)}) +
 {1 \over 2\vf} (\nabla \vf)^2  -
 {\vf \over 4} {\rm tr} (\nabla \sigma \sigma^{-1}
 \nabla \sigma \sigma^{-1}) -
 {\vf^2 \over 4} \sigma_{mn} \,\vf^m_{ij} \,\vf^{nij} \biggr] \, ,
  \label{2.8}
  \end{equation}
 where $\vf^m_{ij} \equiv \partial_i \vf^m_j -
 \partial_j \vf^n_i$. This complicated action can be simplified for some interesting  cases. With a special choice of the
 matrix $\sigma_{mn}$,
 \begin{equation}
 \sigma_{22} = e^{\psi} , \,\,\,\, \sigma_{33} =
 e^{-\psi}, \,\,\,\, \sigma_{23} =
 \sigma_{32} = 0\,,
  \label{2.9}
  \end{equation}
  and the vanishing $\vf_2^m$,
one can derive the simpler effective action:
    \begin{equation}
\cL_{\textrm{c}} = \sqrt{-g^{(2)}} \, \biggl[ \vf R(g^{(2)}) +
        {1 \over 2\vf} (\nabla \vf)^2- {\vf \over 2} \, (\nabla \psi)^2
  -{Q_1^2 \over 2\vf^2} \, e^{-\psi}
   \biggr] \, ,
 \label{2.10a}
    \end{equation}
 where $Q_1$ is the contribution of the integrated out  Abelian gauge
field $\vf_1^m$. This gives us a serious motivation to study
  the DGS models in which scalar potentials are exponential in $\psi$
. Their one-dimensional configuration
can be described by a dynamical system~(\ref{3L}), with the
effective potential $U=e^{g_2\psi}\Phi(\xi)$.
Actually, in pure geometrical case one has $\vf=e^{-\xi/2}$, since
$\dvf/Z=\dxi$ and $Z(\vf)=-\vf/2$. The potential is power-like in
$\vf$ (even taking into account the Weil transformation), and thus
exponential in $\xi$: $\Phi(\xi)=e^{g_1\xi}$. This is the
integrable configuration considered above, but let us first consider
a bit more general potentials with arbitrary $\Phi(\xi)$.

It may be of interest to find how does the general function
$\Phi(\xi)$ break the integrability. The first integral
is obviously $P=2\dpsi-g_2\dxi$ (or, equivalently,
 $P=(g_2-2\cG)\chi$ using $\xi$ as the independent variable).
 If $P$ vanishes, we have the exact expression $\cG=g_2/2$ and then
 applying the MIE procedure easily derive the following special solution:
\begin{equation}
  \psi=\frac{g_2}{2}(\xi-\xi_0),\; \cF(\xi)=\cF_0-\frac{g_{2}^2}{4}(\xi-\xi_0),\; \chi(\xi)=\chi_0-e^{(\cF_0-g_{2}^2\xi_0/4)}\int_{\xi_0}^\xi
    \Phi(\bar{\xi})e^{g_{2}^2\bar{\xi}/4}d\bar{\xi}\,.
\end{equation}

With $P\neq 0$ we return to Eqs.(\ref{eqxi}) and show that the
equation for $\chi$ can be transformed into a simple Hamiltonian
system with `dissipation' vanishing of which transforms it into
completely integrable system. Differentiating $\chi' = -e^{\,\cF +
g_2 \xi} \,\Phi(\xi)$ and using the other equation and integral
$P$,
\begin{equation}
  \psi' = \cG=(g_2-P/\chi)/2 \,,\quad \cF'=-\cG^2=-(g_2-P/\chi)^2/4 \,,
\end{equation}
we derive a somewhat unusual equation for $\chi$ containing a `dissipative' term:
\begin{equation}\label{chionly}
    (\ln\chi')'=\cF'+g_2\psi'+(\ln \Phi)'=\frac{g_{2}^2}{4}-\left(\frac{P}{2\chi}\right)^2+(\ln \Phi)'\,.
\end{equation}

This equation can be rewritten as a Hamiltonian system.  If we denote
$\chi'\equiv \eps e^\rho$ ($\eps=\pm 1$), we find its canonical
 formulation,
\begin{equation}
   \chi'=\eps e^\rho=\frac{\p H}{\p\rho},\quad \rho'= \frac{g_{2}^2}{4}-\left(\frac{P}{2\chi}\right)^2 + (\ln \Phi)'=
    -\frac{\p H}{\p\chi}\,,
\end{equation}
with the $\xi$-dependent (`time' dependent) Hamiltonian,
\begin{equation}
    H=H_0 + \eps e^\rho  -\frac{P^2}{4\chi} - \,\chi g_{2}^2/4 -
    \,\chi (\ln \Phi)'
\end{equation}
 The total derivative of this Hamiltonian on the equations of motion is simply
 \begin{equation}
    \frac{d H}{d\xi} \doteq -\chi (\xi)\,(\ln\Phi(\xi))'' \neq 0\,.
\end{equation}
 For the integrable potential $\Phi=e^{g_1\xi}$, the Hamiltonian is conserved on the equations of motion, $\frac{d H}{d\xi}\doteq 0$, and one can derive the explicit exact solutions. In view of its simplicity, both technical and conceptual, this system may serve as a good laboratory for qualitative and quantitative studies of non-integrable DGS systems.

\subsection{DGS and spherical reduction}

The spherical dimensional reduction implies the following
decomposition of the line element:
\begin{equation}
 ds_D^2 = ds_2^2 + ds_{D-2}^2 =
 g_{ij}\, dx^i\, dx^j \,+ \,
 {\vf}^{2\nu} \, d\Omega_{D-2}^2 \, ,
  \label{2.1}
\end{equation}
where $\nu \equiv (D-2)^{-1}$. It contains only the dilaton, so we
consider the additional ordinary scalar field $\psi$ in the
standard spherically reduced  Einstein gravity:
\begin{equation}
 \label{2.2}
 \cL_{\textrm{s}} =  \sqrt{-g^{(2)}} \,  \biggl[\vf R(g^{(2)}) +
 k_\nu \, \vf^{1-2\nu} +
 {{1-\nu} \over {\vf}} (\nabla \vf)^2 +\vf (\nabla \psi)^2+
  \vf \Phi(\psi)
 \biggr] \,.
\end{equation}

The curvature term, $k_\nu$ rarely supports the symmetry of the
scalar field potential $\Phi(\psi)$. Since we are investigating
the contribution of the scalar mode, we omit the curvature term.
Or, one may consider the vanishing $k_\nu$ as the result of the
trivial cylindrical reduction.
The resulting dynamical system is quite similar to the discussed
above for the non-trivial cylindrical reduction, but it represents
the other variant of the half-exponential potential: now
$U=e^{g_1\xi}\Phi(\psi)$. Indeed, $Z(\vf)=\vf$, so $\dvf/Z=\dxi$
provides $\vf=e^{\xi}$. Taking account of the Weyl transformation and
of the gauge choice we have $g_1=\nu+1$, where normally
$\nu=1,1/2,\ldots,0$, but we suppose that in this DGS model
he parameter $g_1$ may be arbitrary.

In this model, there also exists a linear integral, $P=\dF-g_1\dxi=\chi(\cF'-g_1)+\chi'$, but with the arbitrary potential
$\Phi(\psi)$ this is insufficient for integrability. To find a partial
result, let us combine first and last equations (\ref{eqxi}):
\begin{equation}\label{eq14}
  \cG'=-\frac{\chi'}{\chi}\left(\cG-\frac{dw}{d\psi}\right)\,, \quad
  w(\psi) \equiv \ln{\sqrt{\Phi}}\,.
\end{equation}
 If $P=0$, we can obtain a simple first-order differential equation
 for $y(\psi) \equiv 1/\cG(\psi)$. Recalling that $\chi'/\chi=-\cF'+g_1=\cG^2+g_1$ and $d\xi=d\psi/\cG$,
 we have:
\begin{equation}
  y'(\psi) =\left(1 + g_1\, y^2 \right)
  \left[1 - y \, w'(\psi)\right].
\end{equation}
A nice explicit solution can be obtained if we take $g_1=0$:
$y = e^{-w} \int e^w \,d\psi$.

With non-vanishing $P$,  simple equations with explicit Hamiltonian
interpretation can hardly be derived. For instance, one can
express $\cG=\cG(\chi,\chi')$ from the expression for $P$ and put
it into Eq.~(\ref{eq14}). Then, again using the relation $\psi'=\cG$ one obtains a rather cumbersome second-order
differential equation for $\chi(\psi)$ with arbitrary $\Phi(\psi)$.
 Thus the system with the dilaton half-exponential
potential looks more complicated than that with the scalar
half-exponential potential.

\subsection{DSG and spherical reduction}

Finally we focus on the DSG model -- a specific model of the DGS
class, which is met in affine generalizations of
gravity. The simplest case of DSG implies just one scalar mode, a
scalaron $\psi$, with the Lagrangian of the form \cite{ATFp}:
\begin{equation}
 \label{Lscalaron}
 \cL_{\textrm{DSG}} =  \sqrt{-g^{(2)}} \,  \biggl[\vf R(g^{(2)}) +
 k_\nu \, \vf^{1-2\nu} +
 {{1-\nu} \over {\vf}} (\nabla \vf)^2 +\frac{1}{\vf}(\nabla \psi)^2+
 \Phi(\psi^2/\vf^2)
 \biggr] \,.
\end{equation}

Compared to the ordinary scalar field, it has an abnormal
coupling of the kinetic term to gravity. Now the dependence of $\xi$-variable on the dilaton is  $\vf=\sqrt{2\xi}$, since $Z=1/\vf$ instead of the standard $Z \sim \vf$. In the gauge $s=1/Z$, after the Weyl
transformation with the factor $\Omega=\vf^{\nu-1}$, we then
obtain the effective power-like potential $U(\xi,\psi)=\Omega
Z\Phi=\xi^{\nu/2-1}\Phi(\psi^2/\xi)$. The function
$\Phi(\psi^2/\xi)$ is actually specified in the DSG model, but it
depends on the space-time dimension, and for $D>5$ it cannot be
given explicitly \cite{Filippov:2010bs}. So it is much more
convenient to treat it as an arbitrary function. For the ordinary
scalar field potential, the dilaton field usually enters just as a
factor. For the DSG, in contrast, the mixing of the dilaton and
scalaron is essential.

Again, we omit the curvature term in the potential since it does
not correspond to the inner symmetry of the DSG model. The pure
DSG potential, $U$, satisfies the condition  of the linear symmetry (\ref{Ucond}), even in case of arbitrary $\Phi(\psi^2/\xi)$, if we take
$\lambda=2b_3$, $c_2=-\nu b_3$ with vanishing all remaining parameters.
Then the bilinear integral~(\ref{PDGS}) reads as
\begin{equation}\label{PDSG}
    P=\xi\dF-\frac{\nu}{2}\dxi+\psi\dpsi.
\end{equation}
Unfortunately, to find the other integrals or explicit solutions
we have to require too much: $P$ and $\nu$ must vanish and the
function $\Phi$ has to be specified. Instead, we may use another
approach, the MIE, introduced in~(\ref{MIE}). It suits the DSG
system very well and allows to derive a power-like solution, as we will
see now.

We first choose  $\cG=C_0\xi^{-1/2}$ as a trial function and adjust
the parameters $\xi_0, \psi_0,\chi_0$ to cancel the contribution
of the constant terms. Then we find that
\begin{equation}
    \psi=\psi_0+\cI\{\cG\}=\psi_0+\frac{C_0}{2}\left(\xi^{1/2}-\xi_{0}^{1/2}\right)=\frac{C_0}{2}\xi^{1/2}.\quad
\end{equation}
Actually, we can keep the constant term in the function $\cF$ and find
\begin{equation}
   \cF=\cF_0+\cI\{\cG^2\}=\cF_0-\frac{C_0^2}{4}\ln\xi/\xi_0=\cF_1-\frac{C_0^2}{4}\ln\xi,
\end{equation}
with the new parameter $\cF_1=\cF_0+\frac{C_0^2}{4}\ln\xi_0$.

Next, we see that, fortunately, $\Phi(\psi^2/\xi)= \Phi(C_0^2/4)$
is constant. Therefore we may continue applying the MIE procedure using
an arbitrary DSG potential $\Phi$:
\begin{eqnarray}
\chi(\xi)&=&\chi_0-\frac{e^{\cF_1}\Phi(C_0^2/4)}{-C_0^2/4+\nu/2}\left(\xi^{-C_0^2/4+\nu/2}-\xi_0^{-C_0^2/4+\nu/2}\right)=\nn\\
&&\quad\quad=-\frac{e^{\cF_1}\Phi(C_0^2/4)}{-C_0^2/4+\nu/2}\xi^{-C_0^2/4+\nu/2}.\\
\end{eqnarray}
The final integration in the first MIE iteration provides the
improved function $\cG$ that can be used in the next iteration:
\begin{eqnarray}
\chi(\xi)\cG(\xi)&=&\chi_0 C_0\xi_0^{-1/2}-\frac{e^{\cF_1}C_0\Phi'(C_0^2/4)}{-C_0^2/4+\nu/2}\left(\xi^{-C_0^2/4+\nu/2-1/2}-\xi_0^{-C_0^2/4+\nu/2-1/2}\right)=\nn\\
&&\quad\quad=-\frac{e^{\cF_1}C_0\Phi'(C_0^2/4)}{-C_0^2/4+\nu/2-1/2}\xi^{-C_0^2/4+\nu/2-1/2}.\\
\end{eqnarray}
Of course, to avoid singularities, we suppose that $\nu\neq C_0^2/2+1$,  $\nu\neq C_0^2/2$.

We see that the new function $\cG$ has the same form
$\cG(\xi)=C_1\xi^{-1/2}$, where $C_1$ is a new constant,
\begin{equation}
   C_1=\quad\frac{C_0\Phi'(C_0^2/4)[2\nu-C_0^2]}{2\Phi(C_0^2/4)[2(\nu-1)-C_0^2]}.
\end{equation}
If we choose $C_0$ satisfying the equation $C_1=C_0$, the above expressions
will give an exact solution. It can be useful for the investigation
of the trajectories in vicinity of some singular points.

\section{Conclusion and outlook}

In a set of previous works on the topic of affine generalization
of gravity we discussed the general properties of the theory and
derived a very useful DSG interpretation. Here we tried to advance
toward investigation of the dynamics provided by the theory.
Thus the specific properties of the DSG dynamical system were
revealed. Compared to the closely related but different DGS models,
the DSG configuration appeared to be one of the most favorable for the
analysis, possessing a linear inner symmetry and being well suited
for the iterative MIE procedure.

We suppose that this is not an accident but comes from the
fundamental geometric  origin of the affine gravity, the same as
of the DGS models obtained by various dimensional reductions.
In fact, such relations are usually  hidden and can hardly  be
explicitly demonstrated on the level of \emph{Lagrangians}. So,
for the next step we will try to choose a somewhat different approach
 that would allow to establish a closer relation between the global (topological) properties of the \emph{set of all solutions} and the inner  symmetries of the DGS and DSG models.

In conclusion,  we wish to emphasize that the methods discussed above can be applied outside the context of the affine generalization of
gravity and can provide, as we hope, useful tools for a wide class of
models with scalar fields.

  \renewcommand{\theequation}{A-\arabic{equation}}
  \setcounter{equation}{0}  
  \section*{APPENDIX A: Dynamical systems with linear symmetries}  

Consider a vanishing Hamiltonian for $n$ dynamical
variables $q^i$, which is bilinear in momenta:
\begin{equation}\label{HA1}
    H=A^{ij}(q)p_i p_j+U(q)=0.
\end{equation}
Let us introduce  in the coordinate space an affine transformation
 that defines the vector field
\begin{equation}\label{affinetrans}
  v^i=B^{i}_{j}q^j+c^i,\quad \mbox{where}\quad B^{i}_j,\;c^i=\mathrm{const}.
\end{equation}
For simplicity we will denote the vectors like $\{q^i\}$ by the
bold font, $\qub$. The Euclidean scalar product, $\pb\vb\equiv
(\pb,\vb)=p_i v^i$, is a so-called Hamiltonian of the vector field
$\vb$. Its Poisson bracket with the original Hamiltonian $H$ obviously is
\begin{equation}\label{Poisson0}
    \{H,\pb\vb\}=2(\pb,BA\pb)-\p_v H,
\end{equation}
where $\p_v=v^i\p/\p q^i$ is a derivative along the vector field
$\vb$.

Consider the linear vector fields the matrices of which satisfy
the conditions:
\begin{equation}\label{pBAp}
  (\pb,BA\pb)=\frac{\lambda}{2}(\pb,A\pb)\quad\Rightarrow\quad \{B,A\}\equiv BA+AB=\lambda A+\bar{B},
\end{equation}
where $\bar{B}$ is an arbitrary antisymmetric matrix. Since $H=(\pb,A\pb)+U=0$, the
above relation makes it possible to express the Poisson bracket~(\ref{Poisson0}) as
\begin{equation}\label{Poisson1}
    \{H,\pb\vb\}=2(\pb,BA\pb)-\p_v H=\lambda(\pb,A\pb)-(\pb,\p_v A\pb)-\p_v U=-\lambda U-(\pb,\p_v A\pb)-\p_v
    U.
\end{equation}
It follows that the Poisson bracket of the two Hamiltonians, $H$ and
$\pb\vb$,  vanishes when
\begin{equation}\label{IntCond}
    \{B,A(q)\}=\lambda A(q)+\bar{B}(q),\quad \p_v A^{ij}(q)=0,\quad \p_v U(q) =-\lambda U(q).
\end{equation}
So, in case of the corresponding symmetries of the kinetic matrix
and the potential term, dynamical systems~(\ref{HA1}) admit the
first integrals, which are linear in momenta and coordinates
and are the Hamiltonians of the linear vector fields
$\vb=B\qub+\cb$. Their algebra reads
\begin{equation}\label{PAlgebra}
\{\pb\vb,\pb\vb'\}=\pb\vb'',\quad\mbox{where}\quad \vb''=[B',B]\qub+(B'\cb-B\cb'),
\end{equation}
where the square brackets denote the matrix commutator.

One can easily resolve the conditions~(\ref{IntCond}) for the
subgroup of transformations, $\vb=\lambda\qub+\cb$, which represent
dilatations and  translations:
\begin{eqnarray}
&&A^{ij}(\qub)=\Phi_{ij}(\qub_{\bot}),\\
  && \lambda=0: \; U(\qub)=\Phi_1(\qub_{\bot}),\quad \qub_{\bot}=|\cb|^2\qub-(\qub,\cb)\cb; \label{Ub0}\\
  && \lambda\neq 0:  \;U(\qub)=|\vb|^{-2}\Phi_2(\vb/|\vb|).\label{Ub}
\end{eqnarray}
Here $\Phi_i,\Phi_{ij}$ are the arbitrary functions of their arguments,
which are now  not independent: $\qub_{\bot}$ is orthogonal to
$\cb$, and $\vb/|\vb|$ has unit length.

  \renewcommand{\theequation}{B-\arabic{equation}}
  \setcounter{equation}{0}  
  \section*{APPENDIX B: Linear integrals in DGS}  

Now let us return to the system~(\ref{3EqL}) and try to
construct integrals that are linear in momenta. For simplicity, we
 consider only the symmetries corresponding to the constant
vector fields, $\vb=\cb$. Then the first relation in the
conditions~(\ref{IntCond}) is automatically fulfilled for
$\lambda=0$, since $B$ is now zero matrix.

The remaining conditions can be satisfied in the following way.
Consider first the dependence of the kinetic matrix only on the
dilaton, $A_{ij}=A_{ij}(\vf)$, where the $h$-dependence was
removed by  using  the new variable $\dF=\dhh/h$. The
coordinate vector is now $\qub=(F,\vf,\psi_1,\ldots,\psi_N)$. Thus
a possible ansatz has the form
$\cb=(\alpha,0,\beta^1,\ldots,\beta^N)$, where $\alpha, \beta^i$
are constants. To compactify the lengthy formulas let $\{\beta^i\}$ and $\{\psi^i\}$ be treated as vectors $\betav$, $\psiv$ in
$\mathbb{R}^N$.

All such constant fields, $\cb$, obviously satisfy $\p_c
A_{jk}(q)=c^i\p A_{jk}(\vf)/\p q^i=0$.  The solution for the
potential term for such vector field is already obtained
in~(\ref{Ub0}). Construct first the vector $\qub_{\bot}$
--- the part of $\qub$ that is orthogonal to $\cb$:
\begin{equation}\label{qubot}
q_{\bot}^F=|\betav|^2 F-\alpha\betav\psiv,\; q_{\bot}^\vf=(\alpha^2+|\betav|^2)\vf,\;
q_{\bot}^i=(\alpha^2+|\betav|^2)\psi^i-(\alpha F+\betav\psiv)\beta^i,\;i=1..N\,.
\end{equation}
The potential can be an arbitrary function of $\qub_{\bot}$
components:  $\Phi(F;\vf;\psi_1,\ldots,\psi_N)=\Phi(\qub_{\bot})$.
Yet, in the actual gravitational potential, the metric enters as $e^F$. So we can take the component $q_{\bot}^F$ as the argument of the exponent function. The metric can be excluded from the remaining components $q_{\bot}^i$ if we use the scalar products $q_{\bot}^i \beta^{(m)}_{\bot i}=(\alpha^2+|\betav|^2)\betav^{(m)}_{\bot}\psiv$ with $N\!-\!1$ linearly independent vectors orthogonal to $\betav$: $\betav \betav^{(m)}_{\bot}=0$, $m=1..N\!-\!1$. Then we have the following expression for the potential:
\begin{equation}\label{Uvf}
e^F U(\vf;\{\psi\})=e^{F-\psiv\betav\alpha/|\betav|^2}X(\vf;\{\betav^{(m)}_{\bot}\psiv\}),\quad m=1..N\!-\!1,
\end{equation}
where $X$ is an arbitrary function of $N$ arguments.
The corresponding linear integral is
\begin{equation}
  \pb\cb=\alpha\dvf+2Z_{ij}(\vf)\beta^i\dpsi^j=P\,.
\end{equation}

In the same way we can add the dependence of $Z_{ij}$ on
$\psi_{i_k}$, $k=1,..,K$.  Then we should consider the constant vector $\betav$
with vanishing those $i_k$-th coordinates, actually belonging to $\mathbb{R}^{N\!-\!K}$. And the potential satisfying our conditions will be
\begin{equation}\label{Uvfpsi}
  e^F U(\vf;\psi)=e^{ F-\psiv\betav\alpha/|\betav|^2}X(\vf;\{\psi_{i_k}\},\{\betav^{(m)}_{\bot}\psiv\}),\quad k=1..K,\; m=1..N\!-\!K\!-\!1,
\end{equation}
where now $\betav^{(m)}_{\bot}$ are $N\!-\!K\!-\!1$ arbitrary linearly
independent  constant vectors, belonging to the the orthogonal
complement to $\betav$ in $\mathbb{R}^{N\!-\!K}$ (they also have
vanishing $i_k$-th components). The first integral will be
\begin{equation}
  \pb\cb=\alpha\dvf+2 Z_{ij}(\vf;\{\psi_{i_k}\})\beta^i\dpsi^j=P\,.
\end{equation}

\bigskip
\bigskip
{\bf Acknowledgment.}\\
 This work was supported in part by the Russian Foundation for Basic
 Research: \\ Grant No. 11-02-01335-a and Grant No.
 11-02-12232-ofi-M-2011.

\end{document}